\documentclass[aps,prc,notitlepage,amsmath,amssymb,showpacs,superscriptaddress,twocolumn,nofootinbib]{revtex4-2}

\usepackage{graphics}
\usepackage{graphicx}
\usepackage{color}
\usepackage{hyperref}
\usepackage{bm}
\usepackage{calligra}
\usepackage[T1]{fontenc}
\usepackage{amsmath}
\usepackage{amssymb}
\usepackage{amsfonts}
\usepackage{mathrsfs}  
\usepackage{dsfont}
\usepackage{footmisc}

\usepackage[utf8]{inputenc}

\usepackage{lipsum}
\usepackage{textcomp}
\usepackage{xcolor}

\usepackage{xspace}
\newcommand{\trento}{%
T\raisebox{-.5ex}{R}ENTo\xspace%
}

\begin{document}

\title{Assessing the ultracentral flow puzzle in hydrodynamic modeling of heavy-ion collisions}

\author{A.V.~Giannini}
\email{giannini@ifi.unicamp.br}
\affiliation{Instituto de F\'isica Gleb Wataghin, Universidade Estadual de Campinas, Rua S\'ergio Buarque de Holanda 777, 13083-859 S\~ao Paulo, Brazil}

\author{M. N. Ferreira}
\email{mnarciso@ifi.unicamp.br}
\affiliation{Instituto de F\'isica Gleb Wataghin, Universidade Estadual de Campinas, Rua S\'ergio Buarque de Holanda 777, 13083-859 S\~ao Paulo, Brazil}

\author{M.~Hippert}
\email{hippert@illinois.edu}
\affiliation{Illinois Center for Advanced Studies of the Universe\\ Department of Physics, University of Illinois at Urbana-Champaign, 1110 W. Green St., Urbana IL 61801-3080, USA}

\author{D.D.~Chinellato}
\email{daviddc@g.unicamp.br}
\affiliation{Instituto de F\'isica Gleb Wataghin, Universidade Estadual de Campinas, Rua S\'ergio Buarque de Holanda 777, 13083-859 S\~ao Paulo, Brazil}

\author{G. S. Denicol}
\email{gsdenicol@id.uff.br}
\affiliation{%
	Instituto de F\'isica, Universidade Federal Fluminense,
	Av. Milton Tavares de Souza, Niter\'oi, Brazil, Zip Code: 24210-346,
}%

\author{M.~Luzum}
\email{mluzum@usp.br}
\affiliation{Instituto de F\'{\i}sica, Universidade de  S\~{a}o Paulo,  Rua  do  Mat\~{a}o, 1371,  Butant\~{a},  05508-090,  S\~{a}o  Paulo,  Brazil}

\author{J.~Noronha}
\email{jn0508@illinois.edu}
\affiliation{Illinois Center for Advanced Studies of the Universe\\ Department of Physics, University of Illinois at Urbana-Champaign, 1110 W. Green St., Urbana IL 61801-3080, USA}

\author{T.~Nunes da Silva}
\email{t.j.nunes@ufsc.br}
\affiliation{Departamento de  F\'{\i}sica - Centro de Ci\^encias  F\'{\i}sicas e Matem\'aticas, Universidade Federal de Santa Catarina, Campus Universit\'ario Reitor Jo\~ao David Ferreira Lima, Florian\'opolis 88040-900, Brazil}

\author{J.~Takahashi}
\email{jun@ifi.unicamp.br}
\affiliation{Instituto de F\'isica Gleb Wataghin, Universidade Estadual de Campinas, Rua S\'ergio Buarque de Holanda 777, 13083-859 S\~ao Paulo, Brazil}

\collaboration{{The ExTrEMe Collaboration}\textsuperscript{\hyperlink{extrm}{\S\S}}}

\date{\today}

\begin{abstract} 
An outstanding problem in heavy-ion collisions is the inability for models to accurately describe ultra-central experimental flow data, despite that being precisely the regime where a hydrodynamic description should be most applicable. 
We reassess the status of this puzzle by computing the flow in ultra-central collisions obtained from multiple recent Bayesian models that were tuned to various observables in different collision systems at typical centralities. While central data can now be described with better accuracy than in previous calculations, tension with experimental observation remains 
 and worsens as one goes to ultra-central collisions.  Tuning the model parameters cannot remove this tension without destroying the fit at other centralities.  
As such, new elements are likely needed in the standard modeling of heavy-ion collisions.
\end{abstract}

\maketitle

\begingroup
\setlength\footnotemargin{5pt}
\renewcommand{\footnotelayout}{\hspace{3.5pt}}
\renewcommand*{\thefootnote}{\fnsymbol{footnote}}
\footnotetext[10]{\hypertarget{extrm}{\href{https://sites.ifi.unicamp.br/extreme/}{The ExTrEMe  
Collaboration}} (\emph{``Experiment and Theory in Extreme MattEr''}) 
is a group of researchers focused on  phenomenology of High Energy Heavy Ion Collisions, with special interest in connecting theory with experiments. }
\renewcommand*{\thefootnote}{\arabic{footnote}}
\setcounter{footnote}{0}
\endgroup


\section{Introduction}

High energy heavy-ion collisions are currently understood via complex, 
multistage hybrid hydrodynamic simulations usually composed of 
{\it i}) an initial condition model; 
{\it ii}) a pre-equilibrium phase; 
{\it iii}) a hydrodynamical expansion;
{\it iv}) the conversion from fluid to particles (``particlization'');
{\it v}) final state dynamics.
Each one of these ingredients requires its own model that contains 
various inputs. Recently, the (potentially large) parameter space associated to these 
simulations started to be systematically constrained by means of 
Bayesian parameter estimation. 
Of particular interest are the constraints put on the temperature 
dependence of transport coefficients such as shear and bulk viscosities
that characterize important properties of the quark-gluon plasma --- the 
strongly interacting and deconfined medium produced in heavy-ion collisions --- but are notoriously difficult to calculate from first principles, either numerically~\cite{Meyer:2011gj,Bazavov:2019lgz}
or via perturbative techniques~\cite{Ghiglieri:2018dib,Moore:2020pfu}.

There are several recent comprehensive Bayesian analyses focused on heavy-ion 
(as well as light-heavy ion) collisions~\cite{Bernhard:2019bmu,
JETSCAPE:2020shq, Moreland:2018gsh, Nijs:2020roc, JETSCAPE:2020mzn,
JETSCAPE:2021ehl, Parkkila:2021tqq, Nijs:2021clz, Parkkila:2021yha}.
While each analysis focuses on a particular set of collision 
systems and selected observables, all data considered in these studies come from measurements 
performed at typical centralities (covering at minimum 5\% of the total cross section). Therefore, any constraints produced 
by these studies do not rely on the regime of ultra-central 
collisions, even though there exist data, for example, on flow harmonics 
at the energies considered~\cite{ATLAS:2019peb,CMS:2013bza}. 
This opens up the opportunity to test, for the first time, whether (and how) the 
constraints coming from non-ultra-central collisions affects 
the description of observables measured at ultra-central collisions.

The physics of ultra-central collisions is one where the 
impact parameter nearly vanishes, thus fixing (on average) the 
collision geometry (rotationally invariant for non-deformed nuclei).
Therefore, any quantity must be driven by event-by-event 
fluctuations of the nuclear geometry. This is qualitatively different
from non-central collisions where anisotropic flow develops 
due to stronger pressure gradients along the shorter direction 
of the ``almond'' shaped system formed by the collision 
participants~\cite{Voloshin:2008dg}.
In the latter case, the magnitude of elliptic flow depends crucially on the collision dynamics.   Different models for 
describing the system at extremely early-times (for example a Glauber-type model versus a Color-Glass-Condensate picture \cite{Hirano:2005xf}) predict different spatial eccentricities, which then determine final elliptic flow.  In ultra-central collisions, on the other hand, all flow harmonics are generated only by fluctuations, and specifically are thought to be dominated by initial-state fluctuations driven by quantum fluctuations in nucleon positions within the nucleus before collision.    Further, central collisions achieve the highest temperatures and have the longest lifetime.  Because of this, modern hydrodynamic models should be most reliable in central collisions, and one would expect the best agreement with measured data.

Despite this, for ultra-central events involving 1\% or less of the total number of collisions, for nearly a decade there is a universal inability for hydrodynamic models to simultaneously describe measured flow harmonics $v_n$.   Most notably, either the calculated elliptic flow $v_2$ is larger than measurement, or the predicted triangular flow is too small (or both) \cite{Luzum:2012wu,Roland:2013aaa}.  While the size of the discrepancy varies considerably (even when the problem was first noticed, some calculations approached 10\% agreement at 0--1\% centrality \cite[Fig.~2b]{Luzum:2012wu}), it is striking how universal it is even at a qualitative level, in a regime that naively should be well understood.

Attempts to better understand this puzzle have been made: 
it has been verified that the Monte Carlo Kharzeev–Levin–Nardi and MC-Glauber initial conditions 
generated initial state fluctuations were incompatible with the 
ultra-central data for any value of shear viscosity and that 
repulsive nucleon-nucleon correlations had negligible impact in the 
final anisotropic flow produced by these models~\cite{Shen:2015qta,Qian:2016pau}. 
While the \texttt{magma} model~\cite{Gelis:2019vzt}
is consistent with flow measurements in the $0-1\%$ 
centrality bin~\cite{ATLAS:2019peb} in an initial-state based description, 
this is not true anymore when using it as initial condition to a full 
hydrodynamic simulation~\cite{Snyder:2020rdy}; Ref.~\cite{Snyder:2020rdy} 
also explored alternate initial conditions, such as a modified version of the
\texttt{magma} model, as well as IP-Jazma initial conditions~\cite{Nagle:2018ybc}
(without sub-nucleonic degrees of freedom), which reproduces several features 
present in the IP-Glasma model without using lumpy initial conditions. 
No simultaneous description of $v_n\{2\}$ has been achieved. 
The role of sub-nucleonic fluctuations was explored 
in~\cite{Loizides:2016djv} and did not help solve the issue.
Reference~\cite{Denicol:2014ywa} studied the effect of initial-state nucleon-nucleon correlations, and Ref.~\cite{Rose:2014fba} studied the effect of bulk viscosity, both in the context of the IP-Glasma + hydrodynamics model.
Furthermore,  
it has been pointed out in \cite{Carzon:2020xwp} that including an octupole deformation 
on the lead nucleus can somewhat improve the description of $v_{2}\{2\}-v_{3}\{2\}$, 
but only at the expense of worsening the ratio of triangular flow cumulants, $v_{3}\{4\}-v_{3}\{2\}$. 
Despite all of these efforts, the accurate and simultaneous description of flow harmonics
in ultra-central collisions remains an open problem.

In this work, we reassess for the first time the status of the ultracentral flow puzzle in light of state-of-the-art hybrid hydrodynamic models that have been tuned with extensive model-to-data comparison \cite{Moreland:2018gsh,JETSCAPE:2020mzn,JETSCAPE:2020shq, Nijs:2020roc, Nijs:2020ors, Nijs:2021clz}.   In particular, these models show excellent agreement with data in a range of typical centralities, and represent our best understanding of the collision system to date. This provides a unique opportunity to investigate ultra-central data and learn from the centrality dependence of these models, which is done for the first time in this work. We find that, while central data can now be better described than in previous (pre-Bayesian) calculations, disagreement with experimental data remains 
 and worsens as one moves towards ultra-central collisions. We show that tuning the model parameters of current simulations cannot fix this puzzle without destroying the fit at other centralities.   



\section{Multistage hybrid hydrodynamic simulations}\label{sec:hybridsim}

While modern Bayesian analyses share the basic structure described above,
each analysis is unique due to {\it i})
the specific details of how each stage is implemented and {\it ii}) the 
collision system(s) (as well as the center of mass energy) and observable 
considered in each case. Here we provide a brief summary of the ingredients 
used in Refs.~\cite{Moreland:2018gsh,JETSCAPE:2020mzn,JETSCAPE:2020shq, Nijs:2020roc, Nijs:2020ors, Nijs:2021clz}; we refer the reader 
to these references for a complete description of each simulation.

Due to its flexibility, each analysis employed the (boost invariant) 
\trento parametrization~\cite{Moreland:2014oya} to model the initial condition of their simulation; the notable 
difference being that Refs.~\cite{Moreland:2018gsh,Nijs:2020roc, Nijs:2020ors, Nijs:2021clz} allow for fluctuations of 
nucleon shape via sub-nucleonic degrees of freedom (``hot spots'') while~\cite{JETSCAPE:2020mzn,JETSCAPE:2020shq} considers 
a symmetric nucleon with only a varying radius.
Once the initial energy density is obtained, the system is evolved according to a free streaming prescription until a 
given proper-time $\tau_{\rm fs}$. 
References \cite{Moreland:2018gsh, Nijs:2020roc, Nijs:2020ors, Nijs:2021clz} rely on a centrality independent value for $\tau_{\rm fs}$. 
The JETSCAPE collaboration \cite{JETSCAPE:2020mzn}, on the other hand, allowed $\tau_{\rm fs}$ 
to fluctuate event by event with some power of the mean energy density 
deposited at the transverse plane at the initial time with respect 
to some (arbitrarily defined) scale. The normalization 
and the power to which $\tau_{\rm fs}$ depends on the mean energy 
density were constrained by their Bayesian analysis.  Further, while Refs.~\cite{Moreland:2018gsh, JETSCAPE:2020mzn,JETSCAPE:2020shq} assume a scale-invariant system of massless particles moving at the speed of light, Refs.~\cite{Nijs:2020roc, Nijs:2020ors, Nijs:2021clz} allow for a varying (though still uniform within a system) speed.

At the end of the free-streaming phase, it is assumed that the medium 
can be described by the relativistic viscous hydrodynamics 
equations of motion. All simulations solve second order Israel-Stewart-like hydrodynamic equations ~\cite{IS1976,Israel:1979wp,Denicol:2010xn,Denicol:2012cn}. 
An equation of state is needed in order to close the system of equations; 
all simulations use the the HotQCD~\cite{HotQCD:2014kol} equation of state 
at high temperatures while the low temperature regime is described by a hadron 
resonance gas with different particle content, matched to the hadronic afterburner 
used in each calculation.

\begin{figure}
 	\begin{center}
 		\includegraphics[scale=0.4]{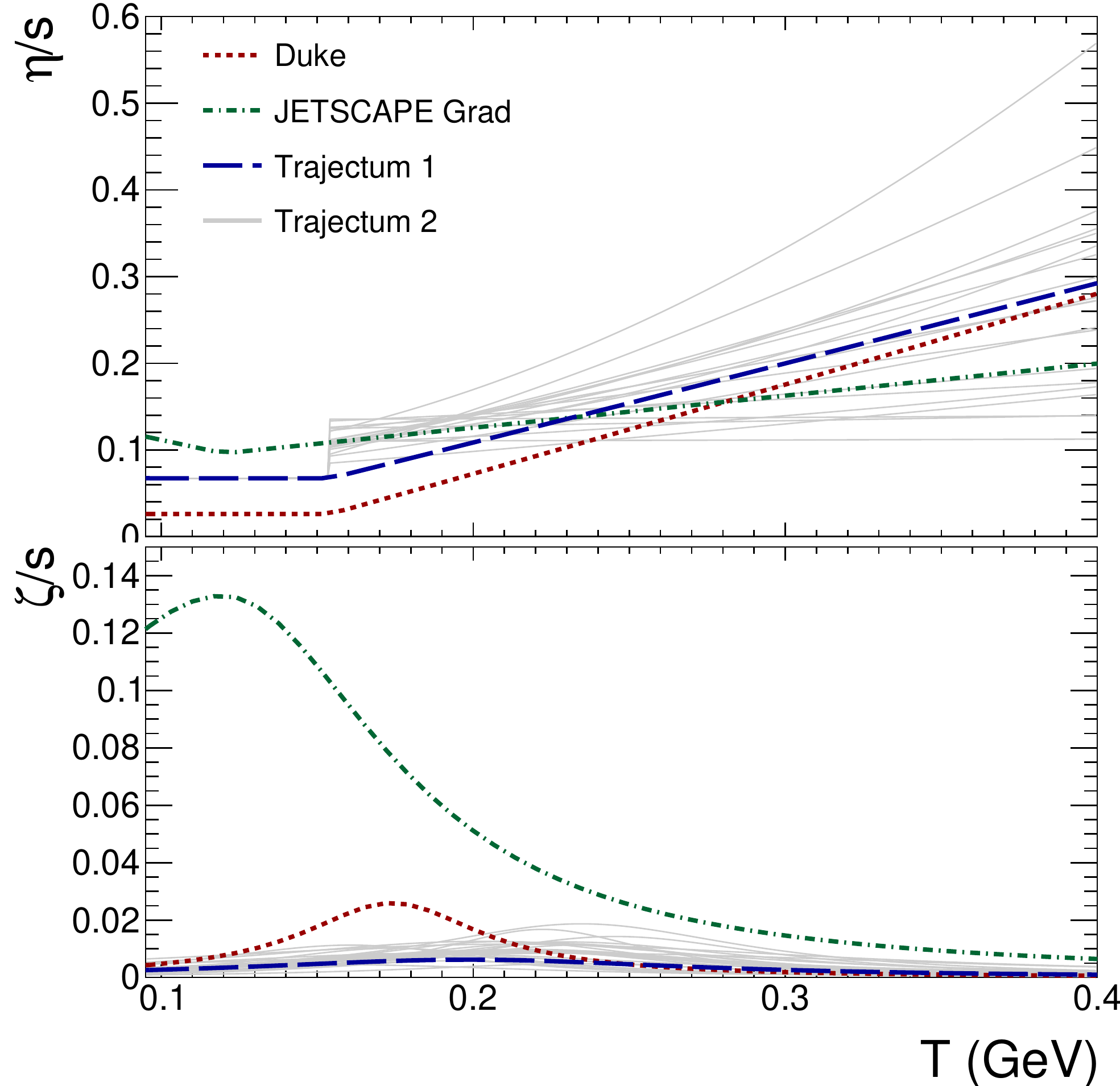}
 	\end{center}
	\caption{Temperature dependence for (top) shear and (bottom) bulk viscosity resulting from different Bayesian constrained models considered in this work. 
	For \emph{Trajectum} 2 we show the resulting temperature dependence for 20 random samples of the full posterior (see text for details).}
	\label{fig:shear_bulk_panel}
\end{figure}

Regarding the temperature dependence of the first-order transport coefficients, each 
calculation uses a similar parametrization for the bulk viscosity: an unnormalized (possibly skewed) Cauchy distribution. 
The functional form assumed for the temperature dependence of the shear viscosity are also similar --- two regions separated by a sudden 
change in slope.  References \cite{JETSCAPE:2020mzn,JETSCAPE:2020shq} assume two linear sections surrounding a variable transition temperature, 
while Refs.~\cite{Moreland:2018gsh, Nijs:2020roc, Nijs:2020ors, Nijs:2021clz} include possible curvature in the high-temperature region. 
It is important to notice, however, that assuming a similar functional form for the shear and bulk viscosities does not imply those transport coefficients will behave in an equally similar way after the parameter estimation is done. The behavior of each transport coefficient is only known after the value of every parameter needed in those parametrizations is determined and the procedure to do so is highly dependent of the specific details of each Bayesian analysis -- for instance, different groups may start with the same functional form for a given transport coefficient but consider different ranges for one or more input parameters. To illustrate this is indeed the case, the resulting shear and bulk viscosity from the different Bayesian analysis mentioned is shown in Fig. \ref{fig:shear_bulk_panel}.

The switch from a fluid description to a particle description is made at an isotherm surface of temperature $T_{\rm sw}$.
Fluid cells are then converted into particle distributions which 
are independently sampled from the switching hypersurface. 

Due to the presence of dissipative corrections in the hydrodynamic phase, 
the distribution function to be sampled deviates from the equilibrium one. 
There are different ways to estimate the out-of-equilibrium correction.
One option is to add a $\delta f$ contribution, which is linear in 
the dissipative currents, to the equilibrium 
distribution, so that the total distribution is $f = f_{\rm eq} + \delta f$. 
Such Ansatz is, however, valid as long as $\delta f \ll f_{\rm eq}$, otherwise 
it may lead to negative distribution functions at higher momenta \cite{Denicol:2009am}.
In order to avoid any possibility of negative probability densities, 
another option is to make use of ``modified'' equilibrium distribution, 
where viscous corrections are taken into account by re-scaling the 
temperature and momenta of the equilibrium distribution by factors that 
depend on out-of-equilibrium quantities~\cite{McNelis:2021acu}.
The JETSCAPE collaboration has explored four different models for correcting 
the equilibrium distribution of hadrons; in this work we follow the setup 
that makes use of a the $\delta f$ correction from the 14-moment approximation, 
identified as ``Grad'' in~\cite{JETSCAPE:2020mzn,JETSCAPE:2020shq}, which gave a better fit to the experimental data studied in that work. 
The other analyses, on the other hand, make use of the Pratt-Torrieri-Bernhard 
modified equilibrium distribution~\cite{Moreland:2018gsh,Bernhard:2018hnz}.

The hadrons sampled from the freeze-out hypersurface are then 
evolved according to  transport equations. 
References \cite{Moreland:2018gsh, Nijs:2021clz} use the UrQMD
afterburner~\cite{Bass:1998ca,Bleicher:1999xi}, while Refs.\ \cite{JETSCAPE:2020mzn,JETSCAPE:2020shq, Nijs:2020roc, Nijs:2020ors} use SMASH in their main analyses.
The JETSCAPE collaboration presented a systematic comparison of results 
with SMASH~\cite{Weil:2016zrk,smash:github} and UrQMD. 
There were no discernible differences in results for the observables 
studied, as long as one matches the low temperature regime of the 
equation of state to the particle content present in each model.

From a systematic data-to-model comparison, one can extract a posterior  distribution, which represents the probability density for the set of model parameters to have a certain value, given the value of measured data.  The maximum of the posterior in this multidimensional parameter space is known as the Maximum a Posteriori (MAP) point, and represents the most probable set of model parameters.  With these parameters, the models have an excellent agreement with the experimental data used in these analyses.   The Duke analysis \cite{Moreland:2018gsh} tuned their model to p-Pb and Pb-Pb data at 5.02 TeV, while JETSCAPE  \cite{JETSCAPE:2020mzn,JETSCAPE:2020shq} performed a simultaneous analysis of 2.76 TeV Pb-Pb collisions and 0.2 TeV Au-Au collisions.  The first \emph{Trajectum} analysis \cite{Nijs:2020roc, Nijs:2020ors} made a simultaneous analysis of 2.76 and 5.02 TeV Pb-Pb collisions and 5.02 TeV p-Pb collisions, while the second \emph{Trajectum} analysis \cite{Nijs:2021clz} used 2.76 and 5.02 TeV Pb-Pb data.

Thus, we see that while each analysis is substantially similar, there are various differences in model as well as the choice of collision systems and observables.  In principle, these differences can significantly impact the final posterior distribution and the predicted values for other observables.
For example, the maximum bulk viscosity $\zeta/s(T)$ from the MAP of Refs.~\cite{JETSCAPE:2020mzn,JETSCAPE:2020shq} is more than a factor 5 higher than any of the other analyses; moreover, the peak of $\zeta/s(T)$ for each state-of-the-art simulation happens at quite different temperatures (see Fig. \ref{fig:shear_bulk_panel}). Therefore, despite the usage of a similar parametrization, the final temperature dependence for the bulk viscosity is different at a qualitative level.

In the next section, we extend the application of these simulation 
chains to flow in ultra-central Pb+Pb collisions, using knowledge obtained from the Bayesian parameter estimation. We assess 
to what extent the Bayesian constraints help us to understand 
the properties of the system formed in these extreme conditions.

\section{Results}\label{sec:results}

Here we apply the simulation chains described in the previous section 
to calculate two-particle flow harmonics in ultra-central Pb+Pb collisions. 
We use 4 different models that we will refer to as Duke \cite{Moreland:2018gsh}, JETSCAPE \cite{JETSCAPE:2020mzn,JETSCAPE:2020shq}, \emph{Trajectum} 1 \cite{Nijs:2020roc, Nijs:2020ors}, and \emph{Trajectum} 2 \cite{Nijs:2021clz}.
We performed simulations using the maximum {\it a posteriori} 
(MAP) parameter values from the respective Bayesian analyses of Duke~\cite[Table III]{Moreland:2018gsh}, JETSCAPE's ``Grad'' seup~\cite[Table II]{JETSCAPE:2020mzn}, and \emph{Trajectum} 1 \cite{Nijs:2020roc, Nijs:2020ors}.  

For \emph{Trajectum} 2, we follow Ref.~\cite{Nijs:2021clz} and use a sampling of the posterior (20 random samples, as illustrated in Fig.\ 3 of~\cite{Nijs:2021clz}, leading to the first-order transport coefficients shown in Fig. \ref{fig:shear_bulk_panel} of the present work), with the average and standard deviation of the results representing the midpoint and uncertainty of the model prediction.  This latter calculation uses more of the information from the Bayesian analysis in order to quantify systematic uncertainty.  Note that there may be nontrivial correlations between parameters, or redundancies.  Different sets of parameters may result in a roughly equivalent description of the data that was used in the Bayesian analysis.  If one uses only the MAP point to predict new observables (in this case, flow observables in different centrality bins), one cannot rule out the possibility that a different combination of parameters could give an almost equal description of the original data, but a quite different prediction for the new observables.  By using the full posterior to predict ultra-central flow, we learn whether it would be possible to re-tune parameters to fit all centralities at the same time since all correlations between the posteriors of each parameter are naturally taken into account.

Here we study only Pb-Pb collisions and limit the application of each model to the center-of-mass energy where 
the respective Bayesian analysis was carried out.  In order to maximize statistics and allow for analysis of extremely central collisions, we select centrality via the total initial energy at $\tau\to 0$ (i.e., the output of \trento). We note that in experiment, as well as the Bayesian analyses cited, centrality is selected via final multiplicity.  While results can generally depend on how centrality is selected \cite{ATLAS:2019peb, Gardim:2019brr}, here we study the basic flow observable $v_n\{2\}$, where differences due to centrality determination are generally much smaller than the discrepancy between theory and experiment~\cite{Nijs:2021clz} --- in the case of \emph{Trajectum} 1, we verified this explicitly by comparing to results with centrality selected via final multiplicity.

\begin{figure}
 	\begin{center}
 		\includegraphics[scale=0.4]{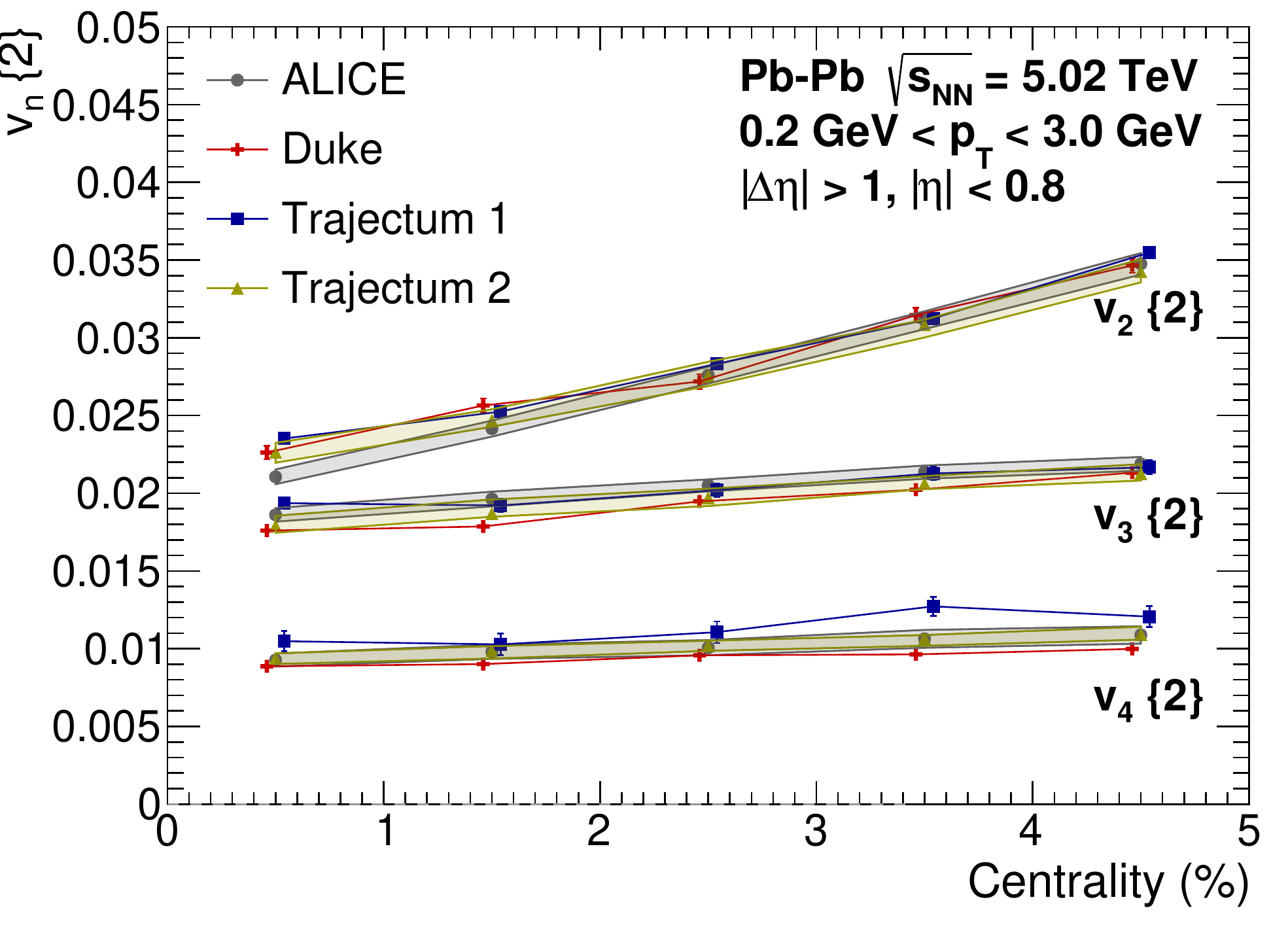}
 	\end{center}
	\caption{Anisotropic flow $v_n\{2\}$ in  1\% centrality bins, with $n=2,3,4$ for Pb+Pb collisions at 5.02 TeV predicted by the MAP parameters from the Duke~\cite{Moreland:2018gsh} and \emph{Trajectum} 1~\cite{Nijs:2020roc, Nijs:2020ors} Bayesian analyses compared to measurements from the ALICE collaboration~\cite{ALICE:2018rtz}.  Also included is the result from \emph{Trajectum} 2 from Ref.~\cite{Nijs:2021clz} for comparison. Duke and \emph{Trajectum} 1 results have been slightly offset horizontally for better visibility.}
	\label{fig:5.02}
\end{figure}

In Figure \ref{fig:5.02}, we show the results of $v_2\{2\}$, $v_3\{2\}$, and $v_4\{2\}$, predicted by the Duke and \emph{Trajectum} 1 setups, compared to measurements from the ALICE Collaboration~\cite{ALICE:2018rtz}.  We note excellent agreement in both cases at centralities above $\approx$ 2\%, both models having been trained from experimental data with the same range in transverse momentum, with the smallest centrality bin representing a combined 0--5\%.  However, the centrality dependence (particularly for $v_2$) appears to deviate from the experimental trend starting at the 1--2\% centrality bin.  We can compare this to the results of \emph{Trajectum} 2 in Ref.~\cite[Fig.~14]{Nijs:2021clz} that we have added to Fig.~\ref{fig:5.02} and which shows the same trend.  

Since the discrepancy between model and data does not exceed 10\% here, one might wonder whether there still remains a puzzle --- even in modern Bayesian analyses, it is difficult to fit all measurements at all centralities at better than $\approx$ 10\% accuracy.  Further, these models were not specifically tuned to reproduce these data, so it might be possible to obtain a better fit (while still retaining a good fit a other centralities) by re-tuning parameters.

To answer this, we first note that the discrepancy with measurement does not appear to be random --- while having a slightly smaller magnitude than in past calculations, it has exactly the same qualitative features (the predicted ratio $v_2$/$v_3$
being too large).  Second, we note that there exist experimental data for more central bins (though only at lower collision energies), and it would be useful to know whether agreement continues to worsen.  Finally, we can partially answer the question of the possibility of retuning parameters to fit ultra-central data by including the systematic uncertainty from the full Bayesian posterior.

\begin{figure}
 	\begin{center}
 		\includegraphics[scale=0.4]{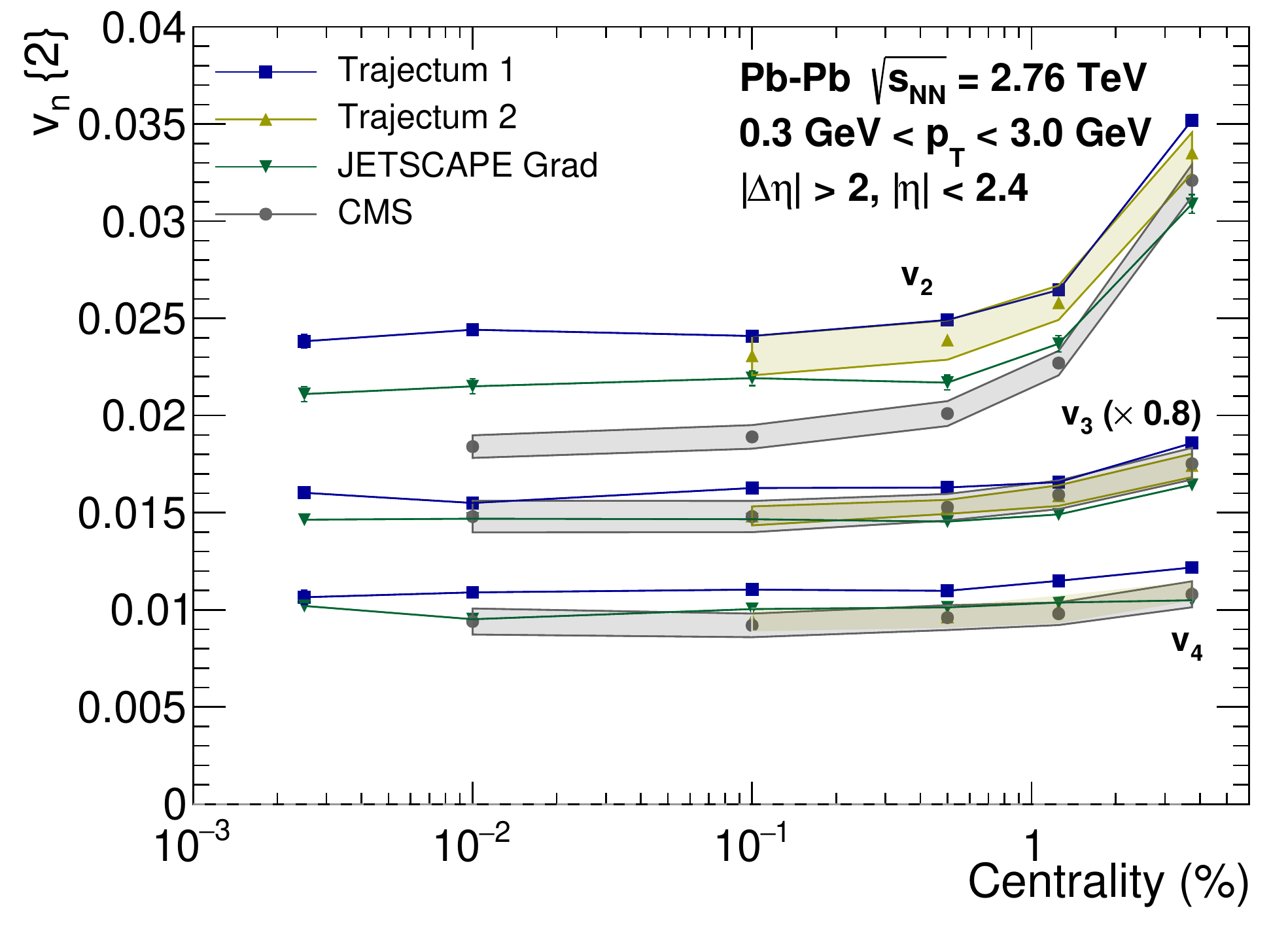}
 	\end{center}
	\caption{Anisotropic flow $v_n\{2\}$ in  central collisions, with $n=2,3,4$ for Pb+Pb collisions at 2.76 TeV predicted by the MAP parameters from the JETSCAPE~\cite{JETSCAPE:2020mzn,JETSCAPE:2020shq} and \emph{Trajectum} 1~\cite{Nijs:2020roc, Nijs:2020ors} Bayesian analyses, as well as the full posterior of  \emph{Trajectum} 2~\cite{Nijs:2021clz}, compared to measurements made by the CMS Collaboration~\cite{CMS:2013bza}. Following the measurement, centrality bins are 0--0.02\%, 0--0.2\%, 0--1\%, 0--2.5\%, and 2.5--5\%, plotted at the midpoint of the bin and on a log scale. An extra bin at 0--0.005\% centrality is added for some models, to verify the trend in centrality. See the text for details.}
	\label{fig:2.76}
\end{figure}

We address all of these in Fig.~\ref{fig:2.76} where we show $v_2\{2\}$, $v_3\{2\}$, and $v_4\{2\}$ measured by the CMS Collaboration~\cite{CMS:2013bza} compared to predictions from the JETSCAPE Bayesian analysis \cite{JETSCAPE:2020mzn, JETSCAPE:2020shq}, as well as \emph{Trajectum} 1~\cite{Nijs:2020roc, Nijs:2020ors} and \emph{Trajectum} 2~\cite{Nijs:2021clz}.  In the latter case, we include a sampling of the full posterior to estimate the Bayesian systematic uncertainty, as described at the beginning of this section.  To better see the individual bins, including ultra-central bins down to 0--0.02\% centrality, we plot centrality on a log scale.

Here we see again that there is good agreement at moderate centralities,\footnote{Each of the original Bayesian analyses were performed using data from the ALICE collaboration, which measures in a different kinematic region than CMS --- in particular, the wider coverage in pseudorapidity ($|\eta|<2.4$ vs.\ $|\eta|<0.8$) allows for a larger rapidity gap to better suppress non-flow ($|\Delta\eta|>2$ vs.\ $|\Delta\eta|>1$).  Likely because of these differences, the agreement of the models at moderate centralities is less perfect here than in comparisons to ALICE data.  See, e.g., the rightmost point in Fig.~\ref{fig:2.76}, representing 2.5--5\% centrality, compared to the same range in Fig.~\ref{fig:5.02}.  We emphasize, however, that this can not explain the different centrality dependence (and therefore the poor agreement in ultra-central collisions).} but a discrepancy appears and indeed continues to get worse as one goes to more and more central collisions. In particular, the experimental elliptic flow continues to decrease with centrality for all measured bins, while the model predictions become almost constant. Even with the included estimate of systematic uncertainty in the \emph{Trajectum} 2 prediction, the agreement becomes poor.

\begin{figure}
 	\begin{center}
 		\includegraphics[scale=0.4]{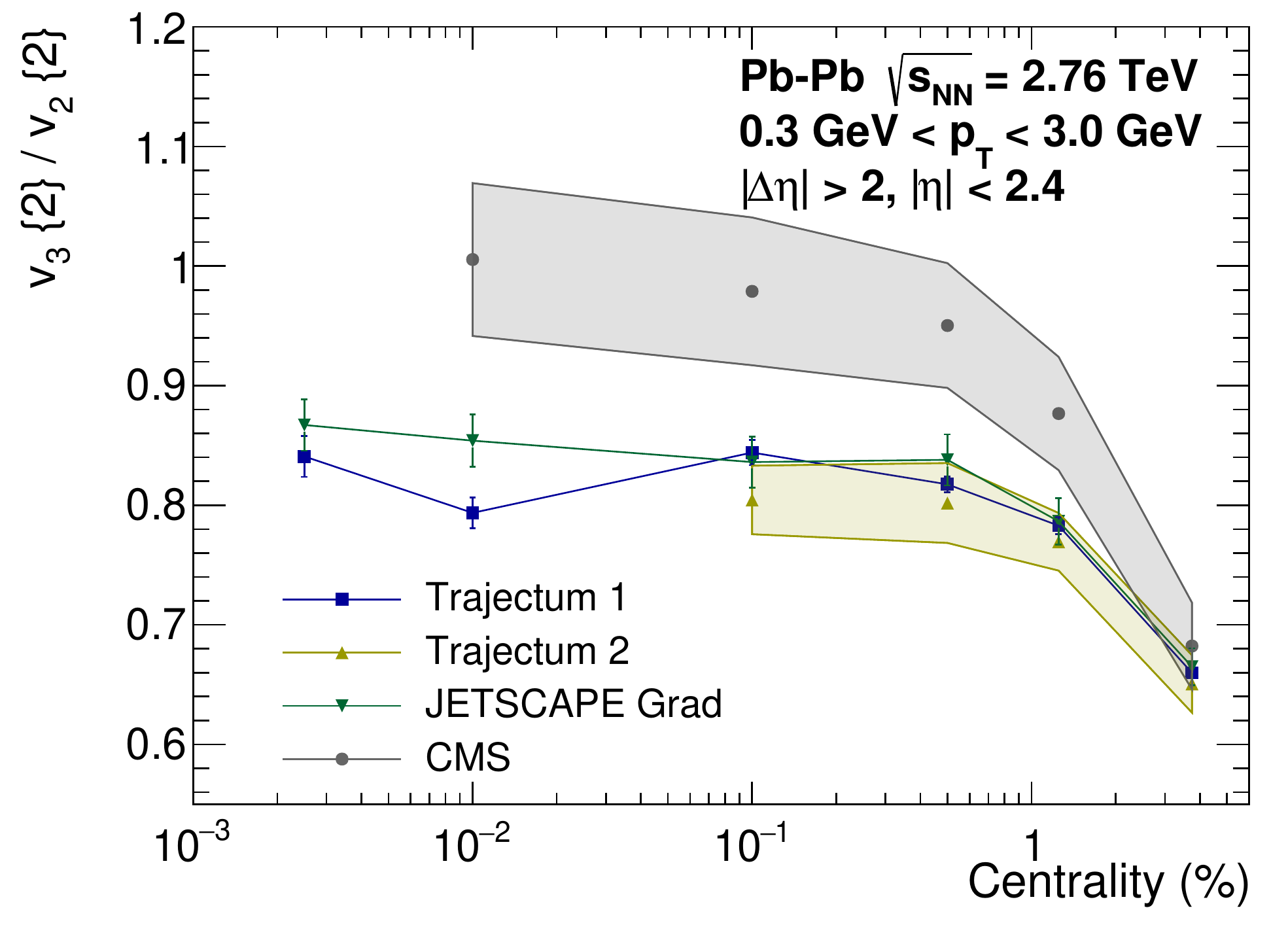}
 	\end{center}
 \caption{Ratio of triangular to elliptic flow $v_3\{2\}/v_2\{2\}$ for Pb+Pb collisions at 2.76 TeV. The CMS data for the numerator and denominator is from~\cite{CMS:2013bza}, with uncertainty in the ratio (over-)estimated by assuming uncorrelated errors.}
 \label{fig:2.76ratio}
\end{figure}

In Fig.~\ref{fig:2.76ratio} we show the ratio $v_3\{2\}/v_2\{2\}$, a simple illustration of the most difficult aspect for simulation models to fit --- relative values of elliptic and triangular flows.  Experimental covariances are not available, so here we estimate the uncertainty on the ratio by assuming uncorrelated uncertainties between $v_2$ and $v_3$.  This is almost certainly an overestimate, but even so, the discrepancy is clear --- the ratio of triangular to elliptic flow in each model simulation reaches an almost constant value, while the experimental data continues to increase with a reduction in centrality, all the way down to 0--0.02\% centrality (and possibly beyond).  In this kinematic range, experimental $v_3$ actually reaches parity with $v_2$, a feature that is not seen in any simulation.

The predictive power of the different models dedicated to describe the several aspects of ultra-relativistic heavy-ion collisions is an important factor to be considered. This is specially important due to the freedom associated with modern Bayesian inference methods that allowed to systematically probe the parameter space of these simulations more efficiently than previous studies. Even though the different state-of-the-art models considered here are being used to extract precision information about quantum chromodynamics (QCD) from heavy-ion data, the extrapolation down to ultra-central collisions is still not well understood, despite being the regime where hydrodynamic based simulations should work best.


If one views the \emph{Trajectum} 2 model as a good overall representation of current Bayesian constrained models when studying ultra-central heavy-ion collisions, our results show that understanding the ultra-central flow puzzle requires going beyond the intuitive picture that the current initial conditions employed are the predominant issue in modern hybrid hydrodynamic simulations. For instance, \emph{Trajectum} 2 results were obtained as the average of 20 unique (random) sets of input parameters for the entire simulation chain.
This is because the plotted band represents the posterior predictive distribution, thus corresponding to many different sets of initial conditions (and its fluctuations), free-streaming times (as well as free-streaming velocities in this particular model) and temperature dependencies for the first-order transport coefficients (as shown in Fig. \ref{fig:shear_bulk_panel}). Thus, we are past the point where one particular set of input parameters was not able to describe ultra-central anisotropic flow data but we might hold out hope to simply find another that could. 

One caveat is that the \emph{Trajectum} analyses used a number of $p_T$-differential observables, combined with only a single model for viscous corrections to the distribution function at particlization.   This results in a narrower posterior parameter distribution than other analyses, e.g., the JETSCAPE results used here.  One could posit that the wider posterior in, e.g., viscosity, could result in a wider posterior predictive distribution for flow in ultra-central collisions, and leaving room for a possibility of obtaining a simultaneous description of flow at all centralities within this wider uncertainty region.

However, since all of the results display a universal centrality dependence of flow in the central region, despite having quite different viscosity, it is more likely that a satisfactory understanding of ultra-central collisions will require modifications to the model for initial conditions.  Currently, geometric fluctuations in models are dominated by the fluctuations in positions of quasi-independent nucleons in the colliding nuclei, with any extra fluctuations in a given model providing a small correction to eccentricities in ultra-central collisions.   It is likely that some new effect will be necessary to understand the true fluctuations in central collisions.

\section{Conclusions}\label{sec:conclusions}

We find that all simulation models that have been tuned with the most recent Bayesian analyses fail to describe the $v_3 / v_2$ ratio in ultra-central collisions.  While flow in moderately-central collisions is described extremely well in some cases (the combined 0--5\% bin being the most central used to train each model), discrepancies appear at centralities near the percent level or lower, and worsen as centrality is decreased (i.e., more central).  These discrepancies follow the universal behavior seen in the past --- either elliptic flow is predicted to be too large, or triangular flow too small.  In particular, the main problem appears to be in the centrality dependence of $v_2$, which has a qualitatively different trend in models compared to experimental data.  We note that, while none of the models included ultra-central data in their respective Bayesian analyses, when we include information from the full Bayesian posterior, this discrepancy remains.  This indicates that it may not be possible to adjust parameters to fit ultra-central data, while maintaining an equivalent fit to the rest of the data. It is thus essential to include ultra-central data in future Bayesian analyses to explicitly check if the current state-of-the-art models can indeed accommodate \textit{all} data simultaneously. Until this is done, one cannot conclude that the ultra-central flow puzzle is solved. 

While the inability of describing ultra-central flow data alone is not a new result, the context in which such disagreement happens is quite different when compared to previous studies and leads to more profound ramifications: results presented in this paper came from multiple state-of-the-art models that went through very resource and time demanding statistical analysis not previously available in hope of improving our knowledge about the characteristics and behavior presented by QCD matter. Exactly because they represent our best understanding of the collision system to date --- including knowledge about initial state fluctuations which are isolated in ultra-central collisions --- it is a striking result that such models still display an overproduction of elliptic flow for the $\lesssim 2\%$ most central collisions. 

The fact that including the systematic uncertainties via samples of the whole posterior distribution produced by the Bayesian analysis did not improve the comparison with experimental data opens up a new path of research to understand the ultra-central flow puzzle: either there is a missing general feature in modern heavy-ion collision hydrodynamic models that prevents a correct description of flow in ultra-central collisions or the current Bayesian analyses need to be augmented by the inclusion of ultra-central data. Therefore, a solution of the ultra-central flow puzzle is bound to shed new light on the theoretical modeling of heavy-ion collisions.

One possible path forward is to improve the description of pre-equilibrium phase in heavy-ion collisions, which currently are described by means of a simple free-streaming model. While a systematic search for what exactly inhibits the understanding of ultra-central flow data is still pending, other possible path is to investigate whether the breaking of the boost invariance in hybrid hydrodynamical simulations in the ultra-central regime plays any role in describing the available data. Studies in this direction are currently ongoing.


\section*{Acknowledgments}
We thank Govert Nijs for making available the \emph{Trajectum} code \cite{trajectum:source} as well as the various parameter values from their Bayesian analyses and the results of Ref.~\cite{Nijs:2021clz} for comparison in Fig.~\ref{fig:5.02}.
This work has been supported by the S\~{a}o Paulo Research Foundation (FAPESP) under projects 
2017/05685-2 (all),
2021/04924-9 (A.V.G.),
2020/12795-1 (M.N.F.),
2016/24029-6, 2018/24720-6, 2021/08465-9 (M.L.),
and 2018/01245-0 (T.N.dS.).
D.D.C., G.S.D, M.L.\ and J.T.\ thank CNPq for financial support. 
J.N. is partially supported by the U.S. Department of Energy, Office of Science, Office for Nuclear Physics under Award No. DE-SC0021301. G.S.D. acknowledges financial support from the Fundação Carlos Chagas Filho de Amparo à Pesquisa do Estado do Rio de Janeiro (FAPERJ), process No. E-26/202.747/2018. 
M.H. was supported in part by the National Science Foundation (NSF) within the framework of the MUSES collaboration, under grant number OAC-2103680.
This research used the computing resources and assistance of the John David
Rogers Computing Center (CCJDR) in the Institute of Physics "Gleb
Wataghin", University of Campinas.

\end{document}